\documentclass[a4paper,10pt]{revtex4}
\usepackage{xcolor}
\definecolor{darkviolet}{rgb}{0.58, 0.0, 0.83}
\definecolor{deepcarminepink}{rgb}{0.94, 0.19, 0.22}
\usepackage[colorlinks,linkcolor={deepcarminepink},citecolor={darkviolet}]{hyperref}
\usepackage{graphicx}
\usepackage{amsmath}
\usepackage{amssymb}
\usepackage{amsfonts}
\usepackage{bm}

\def\be{\begin{align}}
\def\ee{\end{align}}
\def\bea{\begin{eqnarray}}
\def\eea{\end{eqnarray}}

\newcommand{\mc}[1]{\mathcal{#1}}
\newcommand{\f}[2]{\frac{#1}{#2}}

\begin{document}
\title{Axionic extension of the Proca action}
\author{Shahab Shahidi}
\email{s.shahidi@du.ac.ir}
\affiliation{School of Physics, Damghan University, Damghan, 
	41167-36716, Iran.}

\begin{abstract}
	In the context of Cartan theory, we will show that the Proca action can be obtained from the Gauss-Bonnet action for a special choice of the torsion tensor. This in fact equivalent to the special case of the 4th vector Galileon Lagrangian. The theory will then be promoted to contain an axion field. It will be proved that the model admits de Sitter expanding phase with healthy tensor and vector fluctuations. The scalar sector has 4 degrees of freedom, but only one of them remains dynamical in the limit $k\rightarrow\infty$. We will analyze the scalar fluctuations in the small scales limit and obtain the parameter space of the theory in which all the perturbations remain healthy.
\end{abstract}


\maketitle

\section{Introduction}
The universe is in the phase of accelerated expansion \cite{accel}. This expansion however, can not be described by the Einstein's theory of gravity unless one assumes a cosmological constant. Interestingly, such $\Lambda$CDM cosmological model based on the Einstein's theory can be explain observational data very well \cite{LCDM}. However, the addition of a cosmological constant to the theory may cause some phenomenological and theoretical problems \cite{CCproblem}. As a result, it is well-known for decades that one should modify Einstein theory of gravity at least at large scales. There are plenty of gravitational theories, almost all of them explain the accelerated expansion of the universe. Unfortunately, not all of them are healthy and they usually suffer from ghost/tachyon/superluminal/etc. instabilities. As a result, finding a healthy modified gravity theory is an important area of research \cite{modi}.

Modification of gravity can be split in three major categories. The first one is to add some additional fields to the theory which may be considered either as a part of geometry or matter. Examples of such a modification are scalar-tensor \cite{scalten} and vector-tensor theories \cite{vecten}. One of the most interesting scalar-tensor theories is the Galileon theory \cite{galile} which was constructed from the DGP brane-world gravity \cite{DGP} by generalizing the self-interaction of the helicity-0 mode of its graviton. The theory in flat space is written in such a way that under Galilean transformation
$$\phi\rightarrow\phi+b_\mu x^\mu+c,$$
the scalar interactions remain invariant. In the above relation $\phi$‌ is the galileon field and $b_\mu$ and $c$ are constants. Also, the theory has more than second order time derivatives in the action but the resulting equations of motion remain second order. Under the above conditions, there are finite numbers of Lagrangians which we call them ``Galileons". The Galileons have been proved to have a superluminal modes on some special background \cite{super}. Also, they satisfy non-renormalization theorem \cite{nonrenor} which states that at the quantum level, the Galileon interactions do not get renormalized. In fact additional terms always have more derivatives per field than the original Galileon interactions. The covariantization of the Galileons interactions is done in \cite{covgali} where the authors discussed that the higher order time derivatives come back to the equations of motion unless we add some non-minimal scalar interactions with the curvature tensor. These additional terms will then break the Galileon symmetry. The resulting theory is now known as the Horndeski theory \cite{horn}. Many works has been done in the context of Galileon theory \cite{appgali} including cosmology/black hole physics.

An interesting generalization of the Galileon idea is to promote the Galileon interaction to a vector field theory. It has been shown however that such a Galileon interactions can not be constructed for abelian massless vector field and hence the Maxwell's theory can not be generalized in this manner \cite{nogot}. However, considering a Proca Lagrangian, one can construct a finite number of interactions which reduces to the Galileon interactions in the limit $A_\mu=\partial_\mu\phi$, where $A_\mu$‌ is the Proca field and $\phi$‌ is its helicity-0 mode. This generalization of the Proca theory \cite{vecgali} is dubbed vector Galileon. Many works has been done in the context of Galileon theory and also vector-tensor theories, including cosmological implications \cite{genervec} and quantum corrections \cite{quantvec}.

The second category is devoted to generalizing Einstein's theory to contain nonlinear functions of the curvature tensor. Among all of these theories,‌ $f(R)$ gravity \cite{ fR} and Lovelock gravity \cite{lovelock} become more common in the literature. Also, one can assume a non-minimal coupling between matter and geometry and construct theories like $f(R,T)$, $f(R,T,R_{\mu\nu}T^{\mu\nu})$, etc., where $T_{\mu\nu}$‌ is the energy-momentum tensor and $T$ is its trace \cite{fRT}. It is interesting to note that 
the Galileon Lagrangian can be obtained from Lovelock theory by using a suitable boundary term \cite{reunited}. It is then an interesting question that whether the vector Galileon theory can also be obtained from some higher order curvature Lagrangian. This is in fact the scope of the present manuscript.

The last category deals with enriching the geometry of the space-time in order to add some extra degrees of freedom. There are three ways to do this. The first one would be to change the nature of graviton and promote it to a massive tensor degree of freedom \cite{massivereview}. The first attempt goes back to 1939 where Fierz and Pauli construct an action for a massive spin-2 field over flat space \cite{FP}. The Fierz-Pauli theory was recently promoted to the massive gravity action in \cite{dRGT}. Various aspect of the massive gravity theory have been investigated in the literature \cite{massiveapp}.
The second way is to assume that the metric is not anymore compatible with the covariant derivative. This results in the Einstein-Weyl theory which is vastly investigated. The last one is to assume that the torsion tensor is not zero. This results in an Einstein-Cartan theory which assumes that the metric together with the torsion tensor determines the dynamics of gravity \cite{cartan}.

In the Einstein-Cartan theory the connection is no longer symmetric but it is still metric-compatible. The antisymmetric part of the connection is defined as the torsion with the definition
\begin{align}\label{1}
T^\lambda_{~\mu\nu}=\f{1}{2}\left(\Gamma^\lambda_{~\mu\nu}-\Gamma^\lambda_{
~\nu\mu}\right).
\end{align}
With the use of the metric compatibility relation $\nabla_\mu g_{\nu\rho}=0$, one can obtain the connection coefficients in terms of the Christoffel symbol and the torsion as
\begin{align}\label{2}
\Gamma^\lambda_{~\mu\nu}=\left\{\begin{smallmatrix}\lambda\\ 
\mu~\nu\end{smallmatrix}\right\}
+C^\lambda_{\mu\nu},
\end{align}
where we have defined the contortion as
\begin{align}\label{3}
C^\lambda_{~\mu\nu}=T^\lambda_{~\mu\nu}-g^{\lambda\beta}g_{\sigma\mu}T^\sigma_{
~\beta\nu}-g^{\lambda\beta}g_{\sigma\nu}T^\sigma_{~\beta\mu}.
\end{align}
We define the curvature tensor as
\begin{align}\label{4}
K^\lambda_{~\mu\nu\sigma}=\partial_\nu\Gamma^\lambda_{~\mu\sigma}
-\partial_\sigma\Gamma^\lambda_{~\mu\nu}
+\Gamma^\alpha_{~\mu\sigma}\Gamma^\lambda_{~\alpha\nu}-\Gamma^\alpha_{~\mu\nu}
\Gamma^\lambda_{~\alpha\sigma},
\end{align}
with $\Gamma$ is the Cartan connections defined in \eqref{2}. This can be decomposed to the pure metric and pure torsion parts as
\begin{align}\label{5}
K^\lambda_{~\mu\nu\sigma}=R^\lambda_{~\mu\nu\sigma}+C^\lambda_{~\mu\nu\sigma}
,
\end{align}
where
\begin{align}\label{6}
C^\lambda_{~\mu\nu\sigma}&=\nabla_\nu C ^\lambda_{~\mu\sigma}-\nabla_\sigma C 
^\lambda_{~\mu\nu}
+ C ^\alpha_{~\mu\sigma} C ^\lambda_{~\alpha\nu}- C ^\alpha_{~\mu\nu} C 
^\lambda_{~\alpha\sigma}.
\end{align}
One should easily verify that there is only one independent contraction of the curvature tensor, $K^\mu_{~\nu\mu\rho}$ which gives
\begin{align}\label{7}
K_{\mu\nu}=R_{\mu\nu}+\nabla_\lambda 
C^\lambda_{~\mu\nu}+\nabla_\nu C_{\mu},
\end{align}
where $R_{\mu\nu}$ is the Ricci tensor and $C^\alpha=C^{\alpha\beta}_{~~~\beta}$. The other contractions are either zero or can be reduced to the above expression. The scalar curvature can then be obtained by further contraction, with the result
\begin{align}\label{8}
K=R+2\nabla_\lambda 
C^{\lambda}-C^{\alpha}C_{\alpha}
+C_{\alpha\mu\lambda}C^{\alpha\lambda\mu}.
\end{align}

In this paper we are going to investigate the effect of non-zero torsion in the Gauss-Bonnet theory. In four dimensional Riemann space-time, the Gauss-Bonnet Lagrangian is a total derivative due to the Gauss-Bonnet theorem. However, as we will see in this paper, the Gauss-Bonnet Lagrangian in Cartan space is non-zero. In order to write the Gauss-Bonnet action in Einstein-Cartan space-time, one should note that because the symmetry of the curvature tensor $K_{\mu\nu\rho\sigma}$ under the transformation $(\mu\nu)\leftrightarrow(\rho\sigma)$ is lost, one has two independent second order terms in $K_{\mu\nu\rho\sigma}$
\begin{align}\label{9}
K_{\lambda\mu\nu\sigma}K^{\lambda\mu\nu\sigma},\qquad K_{\lambda\mu\nu\sigma}K^{\nu\sigma\lambda\mu}.
\end{align}
Also, because the contracted curvature tensor \eqref{7} is asymmetric, one has two independent second order combinations of the contracted curvature tensor
\begin{align}\label{10}
K_{\mu\nu}K^{\mu\nu},\qquad K_{\mu\nu}K^{\nu\mu}.
\end{align}
In section \ref{sec2}, we will write the most general Gauss-Bonnet action in Cartan space-time. In the special case where only the trace part of the torsion tensor is non-zero, we can recover the second and the 4th vector Galileon Lagrangian. However, we will show that after integration by parts the action is equivalent to the Proca action. This is our first main result in this paper. The possibility of recovering other vector Galileon terms from higher order Lovelock invariants will be investigated elsewhere. For more on the quadratic actions with torsion see \cite{tor}

In section \ref{sec3} we will assume that the axial part of the torsion tensor is also non-zero and construct an axionic extension of the Proca theory (APT). The resulting theory is a two parameter family of tensor-vector-scalar theories, which reduces to the Proca theory theory in the case of vanishing axion field.
 
We then find a de Sitter solution for the theory and show that the tensor fluctuations are always healthy. Demanding that the vector perturbations to be healthy will put a constraint on the parameters of the theory which we will obtain in section \ref{sec5}. We should note that the scalar perturbation in general contains a ghost mode. However, we will show that at deep inside the hirozon limit with $k\rightarrow\infty$, only one scalar mode remains dynamical and assuming that the vector perturbation is healthy, the scalar perturbation will also becomes healthy. We will conclude in section \ref{sec6}.

\section{Vector Galileons via Cartan-Gauss-Bonnet}\label{sec2}
In this section we will consider the effect of torsion tensor in Gauss-Bonnet gravity theory. A more general case was considered in \cite{WCGB}, where the authors consider Weyl-Cartan space in Gauss-Bonnet theory; see also \cite{more}. In this section we will review the results of \cite{WCGB} in a bit different viewpoint.

Let us consider the Gauss-Bonnet action in Cartan space-time. The action functional can be written as
\begin{align}\label{10.1}
S=\int d^4x\sqrt{-g}\bigg[\kappa^2 K-\rho \mathcal{L}_G\bigg],
\end{align}
where $K$ is the curvature scalar in Cartan space-time \eqref{8} and $\rho$ is a dimensionless coupling constant. We have defined the generalized Gauss-Bonnet Lagrangian as
\begin{align}\label{11}
\mathcal{L}_G=\alpha K^{\alpha \beta 
\gamma\delta} K_{\alpha \beta \gamma\delta} + (1-\alpha) K^{\alpha \beta 
\gamma\delta} 
K_{\gamma \delta \alpha \beta} 
- 4\beta K_{\beta\gamma} K^{\beta\gamma} 
-4(1-\beta) K_{\beta\gamma} 
K^{\gamma\beta}+ K^2.
\end{align}
In the above action, $\alpha$ and $\beta$ are two dimensionless constants. The above action then represents a three parameter family of theories in the Cartan space-time. One can easily check that the Lagrangian \eqref{11} reduces to the standard Gauss-Bonnet Lagrangian in the absence of Torsion tensor. In fact the above Lagrangian is the most general Lagrangian which is second order in curvature tensor and reduces to Gauss-Bonnet Lagrangian when $T_{\mu\nu\rho}$ vanishes. For more discussions about this issue see \cite{WCGB}.

The torsion tensor can be decomposed irreducibly into \cite{torsiondecom}
\begin{align}\label{12}
T_{\mu\nu\rho}=\f 2 3 
(t_{\mu\nu\rho}-t_{\mu\rho\nu})+\f13(\hat{Q}_{\nu}g_{\mu\rho}-\hat{Q}_\rho 
g_{\mu\nu})+\epsilon_{\mu\nu\rho\sigma}S^\sigma,
\end{align}
where $\hat{Q}_\mu$ is the trace of the torsion tensor over its first and third indices and $S^\mu$ is an axial vector field. The tensor  
$t_{\mu\nu\rho}$ is antisymmetric with respect to the first two indices and has the 
following properties
\begin{align}\label{13}
t_{\mu\nu\rho}+t_{\nu\rho\mu}+t_{\rho\mu\nu}=0,\quad 
g_{\mu\nu}t^{\mu\nu\rho}=0=g_{\mu\nu}t^{\mu\rho\nu}.
\end{align}
According to the decomposition of the torsion tensor \eqref{12}, one can obtain the contortion tensor as
\begin{align}\label{14}
C_{\rho\mu\nu}=\f 4 3 
(t_{\mu\nu\rho}-t_{\rho\nu\mu})+\f23(\hat{Q}_{\mu}g_{\nu\rho}-\hat{Q}_\rho 
g_{\mu\nu})+\epsilon_{\rho\mu\nu\sigma}S^\sigma.
\end{align}
The effects of the tensor field $t_{\mu\nu\rho}$ is vastly investigated in the context of supergravity theories \cite{supergravity}. In this paper, we will assume that $t_{\mu\nu\rho}$ vanishes for simplicity. In the next section, we will explore the role of the axial vector field $S^\mu$ in the dynamics of the Universe.
But let us for a moment assume that the only non-zero components of the torsion tensor is its trace part $Q_\mu$. So consider a special case
\begin{align}\label{15}
C_{\rho\mu\nu}=Q_{\mu}g_{\nu\rho}-Q_\rho g_{\mu\nu},
\end{align}
where we have defined $Q_\mu=2/3\hat Q_\mu$. After substituting the above expression into equations \eqref{4}-\eqref{8}, the action \eqref{10.1} reduces to
\begin{align}\label{16}
S=\int d^4x\sqrt{-g}&\bigg[\kappa^2R-6\kappa^2Q_\alpha Q^\alpha+8\rho Q_\alpha Q_\beta R^{\alpha\beta}-8\rho\nabla_\alpha Q^\alpha \nabla_\beta Q^\beta\nonumber\\&-8\rho(2\beta-\alpha-1)\nabla_\alpha Q_\beta \nabla^\beta Q^\alpha+8\rho(2\beta-\alpha)\nabla_\alpha Q_\beta \nabla^\alpha Q^\beta\bigg].
\end{align}
Now, by redefining the vector field as $Q_\alpha\rightarrow \sqrt{8\rho} \,Q_\alpha$ and defining the constant $c_2=2\beta-\alpha-1$, one can obtain
\begin{align}\label{17}
S=\int d^4x\sqrt{-g}\bigg[\kappa^2R&-\f12m^2Q_\alpha Q^\alpha+Q_\alpha Q_\beta G^{\alpha\beta}-\mc{L}_4\bigg].
\end{align}
where $\mc{L}_4$ is the 4th vector galileon term defined as \cite{vecgali}
\begin{align}
\mc{L}_4=-\f12Q_\alpha Q^\alpha R+\bigg[\nabla_\alpha Q^\alpha \nabla_\beta Q^\beta+c_2\nabla_\alpha Q_\beta \nabla^\beta Q^\alpha-(1+c_2)\nabla_\alpha Q_\beta \nabla^\alpha Q^\beta\bigg],
\end{align}
and we have defined the vector field mass as $m^2=3\kappa^2/2\rho$. Also, the second and third terms in \eqref{17} can be considered as a second vector Galileon term $\mc{L}_2$, since they do not introduce higher order time derivatives to the action. One should note that we have obtained a special form of the  vector Galileon Lagrangian with $f(Q^2)=Q^2$ (see ref. \cite{vecgali}). In our case, after integrating by parts, one can obtain the Proca theory
\begin{align}\label{17.1}
S=\int d^4x\sqrt{-g}\bigg[\kappa^2R-\f14\hat{Q}_{\mu\nu}\hat{Q}^{\mu\nu}-\f12m_{eff}^2\hat{Q}_\alpha \hat{Q}^\alpha\bigg],
\end{align}
where we have defined $m_{eff}^2=m^2/2(1+c_2)$ and $\hat{Q}_\mu=\sqrt{2(1+c_2)}Q_\mu$.

As a summary, the trace part of the torsion tensor in the Gauss-Bonnet action can produce the vector Galileon Lagrangians $\mc{L}_2$ and $\mc{L}_4$. It will be very interesting to investigate whether higher order Lovelock invariants in Cartan space-time can produce the other vector Galileon terms. This will be done in a separate work. In this paper, we are going to investigate the role of axial part of the torsion tensor $S^\mu$‌ together with the trace part $Q_\mu$ in the theory.
\section{The axionic extension}\label{sec3}
Let us now assume that $S^\mu$ becomes non-zero. The contortion tensor can be written as
\begin{align}\label{18}
C_{\rho\mu\nu}=Q_{\mu}g_{\nu\rho}-Q_\rho g_{\mu\nu}+\epsilon_{\rho\mu\nu\sigma}S^\sigma,
\end{align}
By substituting the above relation into the action \eqref{10.1}, one can obtain a term
$\beta\epsilon^{\alpha\beta\gamma\delta}Q_{\alpha\beta}S_{\gamma\delta}$
where we have defined the strength tensors as $Q_{\mu\nu}=\nabla_\mu Q_\nu-\nabla_\nu Q_\mu,\nonumber$ and likewise for $S_{\mu\nu}$. This term is very similar to the axion interaction term in QCD \cite{axion}. In fact if the axial vector is somehow proportional to the vector field $Q_\mu$, this terms is exactly the axion interaction term. However, this new term is a total derivative and vanished from the action. 

In this paper, we are going to adopt a procedure to keep this term in the action. This will result in an axionic extension of the Proca theory. To do this, we assume that the axial vector field $S_\mu$ is related to the vector field $Q_\mu$ with an axial scalar field $\phi$ which will be the axion field. Note that $\phi$ should be dimensionless. Also, in order to keep the axionic interaction term, one should promote the constant $\beta$ in the Gauss-Bonnet Lagrangian \eqref{11}, to a dynamical scalar field. A straightforward assumption is $\beta=\beta_2(\phi^2)$ (note that $\beta_2$ should be a scalar field). With these in hand, the contortion tensor can be written as
\begin{align}\label{15.1}
C_{\rho\mu\nu}=Q_{\mu}g_{\nu\rho}-Q_\rho g_{\mu\nu}+\beta_1\epsilon_{\rho\mu\nu\sigma} Q^\sigma,
\end{align}
where $\beta_1=\beta_1(\phi)$ is an arbitrary pseudo scalar function constructed from the axion field.
and the Guass-Bonnet Lagrangian is promoted to
\begin{align}\label{11.1}
\mathcal{L}_G=\alpha K^{\alpha \beta 
	\gamma\delta} K_{\alpha \beta \gamma\delta} + (1-\alpha) K^{\alpha \beta 
	\gamma\delta} 
K_{\gamma \delta \alpha \beta} 
- 4\beta_2 K_{\beta\gamma} K^{\beta\gamma} 
-4(1-\beta_2) K_{\beta\gamma} 
K^{\gamma\beta}+ K^2.
\end{align}
As we will see in the following, the axion field acquires a kinetic term from the Gauss-Bonnet Lagrangian and so it becomes dynamical in this theory. We will then define the action of ``axionic extension of the Proca theory" (APT) as
\begin{align}\label{10.2}
S=\int d^4x\sqrt{-g}\bigg[\kappa^2 K-\rho \mathcal{L}_G-\f12\kappa^2\nabla_\alpha\phi\nabla^\alpha\phi-V(\phi)\bigg],
\end{align}
where we have added a potential and a kinetic terms for the axion field for completeness. Upon substituting the contortion tensor \eqref{15.1} to equations \eqref{4}-\eqref{8}, the APT action \eqref{10.2} will be expanded as
\begin{align}\label{act}
S=&\int d^4x\sqrt{-g}\bigg[\kappa^2 R-\f12m^2Q^2+\f12m^2Q^2\beta_1^2+\f{m^2}{4\kappa^2}Q^4\beta_1^2+\f34Q^2\nabla_\alpha\beta_1\nabla^\alpha\beta_1-\f12\kappa^2\nabla_\alpha\phi\nabla^\alpha\phi\nonumber\\&-\f14\beta_1^2 R_{\alpha\beta}Q^\alpha Q^\beta-\f34\beta_1^2 Q_\alpha\Box Q^\alpha-\f14 Q_{\mu\nu}Q^{\mu\nu}-\f{1}{8\alpha}(4\beta_1^2\beta_2-4\beta_2-\alpha\beta_1^2)Q_{\mu\nu}Q^{\mu\nu}\nonumber\\&-\f{1}{8\alpha}(4\beta_2-\alpha)A_{\mu\nu}A^{\mu\nu}+\f{1}{4\alpha}\beta_1(4\beta_2-\alpha)A_{\mu\nu}Q^{\mu\nu}+\f{m}{2\sqrt{3}\kappa}\big[\beta_1 Q^\beta\nabla_\beta\beta_1+3\beta_1^2\nabla_\beta Q^\beta\big]Q^2\nonumber\\&-\f{1}{8\alpha}(1+2\alpha-4\beta_2)\epsilon_{\alpha\beta\gamma\delta}\big(\beta_1 Q^{\alpha\beta}-A^{\alpha\beta}\big)Q^{\gamma\delta}-V(\phi) \bigg],
\end{align}
where we have defined
$$A_{\mu\nu}=Q_\mu\nabla_\nu\beta_1-Q_\nu\nabla_\mu\beta_1,\qquad m=\f{\sqrt{3}}{2}\f{\kappa}{\sqrt{\alpha\rho}},$$
with the condition $\alpha\rho>0$. One should note that in the potential term $V(\phi)$ the even powers of the axion field $\phi$ should be considered. One should note that in the case $\phi=0$‌ and $\beta_1(0)=0=\beta_2(0)$, the above action  reduces to \eqref{17.1} with $\beta=0$‌. Note that the parity violating term becomes total derivative in the case of constant axion field, $\phi=const.$

In order to obtain the field equations of the APT theory, one should vary the action \eqref{act} with respect to the metric $g_{\mu\nu}$, the axion filed $\phi$‌ and the vector field $Q_\mu$. The metric field equation can be obtained as
\begin{align}
&\kappa^2 G_{\alpha\beta}-\f12 \beta_1^2 Q^\nu Q_{(\alpha}R_{\beta)\nu}+\f12 V g_{\alpha\beta}+\f{m^2}{4}\left(1-\beta_1^2\right)\left(Q^2 g_{\alpha\beta}-2 Q_\alpha Q_\beta\right)\nonumber\\
&-\f{m^2}{8\kappa^2}\left(Q^2 g_{\alpha\beta}-4 Q_\alpha Q_\beta\right)\beta_1^2 Q^2+ \f{\sqrt{3}m}{12\kappa}\beta_1\big[2\left(3\beta_1\nabla_\nu Q^\nu+ Q^\nu \nabla_\nu \beta_1\right)Q_\alpha Q_\beta\nonumber\\
&-2\left(6\beta_1 Q_{(\alpha}\nabla_{\beta)}Q_\nu+5 Q_\nu Q_{(\alpha}\nabla_{\beta)}\beta_1 \right)Q^\nu+ \left(6\beta_1 \nabla_\mu Q_\nu +5 Q_\nu \nabla_\mu \beta_1\right)Q^\mu Q^\nu g_{\alpha\beta}\big]\nonumber\\&-\beta_1m^2 Q_{(\alpha}\Box Q_{\beta)}+\nabla_\nu \left(\beta_1^2 Q^\nu \nabla_{(\alpha}Q_{\beta)}\right)+\f12 \beta_1 Q_{(\alpha}\nabla^\nu \nabla_{\beta)}\left(\beta_1 Q_\nu\right)-Q_{(\alpha} \nabla^\nu \left(\beta_1^2 \nabla_{\beta)}Q_\nu\right)\nonumber\\&+Q^2\nabla_\alpha \beta_1 \nabla_\beta\beta_1 - Q_\alpha Q_\beta \beta_1 \Box \beta_1+ 2\beta_1 \left[A_{\nu(\alpha}\nabla_{\beta)}Q^\nu+\nabla^\nu \beta_1 Q_{(\alpha}Q_{\beta)\nu}\right]-\f14 \beta_1^2 Q_{\nu(\alpha}\nabla_{\beta)}Q^\nu\nonumber\\
&+\f{1}{8\alpha}\left(2\beta_2\beta_1^2-2\beta_2-\alpha\right)\left(4Q_{\alpha \nu}Q_\beta^{~\nu}+ Q_{\mu \nu}Q^{\mu\nu} g_{\alpha\beta}\right)-\f2\alpha \beta_1\beta_2 Q_{\nu(\alpha}A_{\beta)}^{~~\nu}-\f1\alpha \beta_2 A_{\alpha\nu}A_{\beta}^{~\nu}
\nonumber\\
&-\f18 \big[\nabla_\nu\left(\beta_1^2 Q^\nu \nabla_\mu Q^\mu\right)+Q^\mu \nabla_\mu \nabla_\nu\left(\beta_1^2 Q^\nu\right)-Q^\mu Q^\nu \nabla_\mu\beta_1 \nabla_\nu \beta_1+4\beta_1^2 \nabla_\mu Q_\nu \nabla^\mu Q^\nu\nonumber\\
&+4\nabla_\nu\beta_1 \nabla^\nu\left(\beta_1 Q^2\right)\big]g_{\alpha\beta}+\f{1}{4\alpha}\beta_2 \left(A^{\mu\nu}-2\beta_1 Q^{\mu\nu}\right)A_{\mu\nu}g_{\alpha\beta}-\f12\nabla_\alpha\phi\nabla_\beta\phi+\f14\nabla_\mu\phi\nabla^\mu\phi g_{\alpha\beta}=0.
\end{align}
The axion field equation of motion can be obtained as
\begin{align}
&\Box\phi+ \f{m^2}{2\kappa^2} \beta_1\beta_1^\prime \left(2\kappa^2+ Q^2\right) Q^2 -\f12 \beta_1\beta_1^\prime R^{\alpha\mu}Q_\alpha Q_\mu +\f{m}{2\sqrt{3}\kappa}\beta_1\beta_1^\prime \left(5Q_\alpha \nabla_\mu Q^\mu-2Q^\mu \nabla_\mu Q_\alpha\right)Q^\alpha\nonumber\\
&+\f12\beta_1^\prime \nabla_\mu\left[ \nabla_\alpha \left(\beta_1 Q^\mu \right)- 4  \nabla^\mu \left(\beta_1 Q_\alpha\right)\right]Q^\alpha+\f{1}{\alpha}\big[2\beta_2 \beta_1^\prime\left(\nabla_\mu A^{\alpha\mu}-\phi \nabla_\mu Q^{\alpha\mu}\right)Q_\alpha\nonumber\\
&+\f12\beta_2^\prime A^{\alpha\mu}A_{\alpha\mu}-\beta_2\beta_1^\prime Q^{\alpha\mu}A_{\alpha\mu}+\f12\beta_2^\prime\left(1-\beta_1^2\right)Q^{\alpha\mu}Q_{\alpha\mu}+\f12\beta_1\beta_2^\prime\epsilon_{\alpha\beta\gamma\delta}Q^{\alpha\beta}Q^{\gamma\delta}\big]-V^\prime(\phi)=0,
\end{align}
where prime denotes derivative with respect to the argument. The vector field equation of motion can be written as
\begin{align}
&\nabla_\mu Q^{\mu\alpha}+\f12 \beta_1 \nabla_\mu \left(A^{\mu\alpha}-\beta_1 Q^{\mu\alpha}\right)-\f32 \beta_1 \Box\left(\beta_1 Q^\alpha\right)+\f{m^2}{\kappa^2} \left(Q^2 \beta_1^2 +\kappa^2\beta_1^2-\kappa^2\right)Q^\alpha\nonumber\\
&+\f{2}{\alpha}\big(\nabla_\mu\left(\beta_2 Q^{\alpha\mu}\right)-\beta_1\nabla_\mu\left(\beta_1\beta_2 Q^{\alpha\mu}\right)+\beta_1\nabla_\mu\left(\beta_2 A^{\alpha\mu}\right)\big)+\f{2}{\alpha}\beta_1\epsilon^{\alpha\beta\mu\nu} Q_{\beta\mu}\nabla_\nu\beta_2\nonumber\\
&+\f{\sqrt{3} m}{6\kappa}\beta_1\left(2 Q_\mu A^{\alpha\mu}+6\beta_1 Q^\alpha \nabla_\mu Q^\mu-3 Q^2\nabla^\alpha\beta_1 -6\beta_1 Q^\mu \nabla^\alpha Q_\mu\right)-\f12 \beta_1^2 Q_\mu R^{\alpha\mu}=0.
\end{align}

\section{Cosmology of the APT gravity}
Let us consider the flat FRW Universe as
\begin{align}
ds^2=-dt^2+a^2\left(dx^2+dy^2+dz^2\right),
\end{align}
where $a=a(t)$‌ is the scale factor of the universe with its associated Hubble parameter defined as $H=\dot{a}/a$ where dot denotes time derivative. 
The isotropy and homogeneity
 conditions impose that the vector field only has the temporal component with the form $Q_\mu=(Q(t),0,0,0)$ and the scalar field is only a function of time, $\phi=\phi(t)$.

The only non-zero component of the vector field equation is its temporal component
\begin{align}
\f{m^2}{\kappa^2}  \left(\kappa^2(1-\beta_1^2)+\beta_1^2 Q^2\right)Q&+\f{\sqrt{3}m}{2\kappa}\beta_1 (6H\beta_1-\dot{\beta_1})Q^2+\f32\beta_1^2 (4H^2+\dot{H})Q\nonumber\\-\f92 H\beta_1(Q\dot{\beta_1}+\beta_1 \dot{Q})&-\f32 \beta_1 (2\dot{\beta_1}\dot{Q}+\beta_1\ddot{Q}+Q \ddot{\beta_1})=0.
\end{align}
The scalar field equation reduces to
\begin{align}
 \f{\beta_1\beta_1^\prime}{ 2\kappa^2} \big[m^2&\left(Q^2-2\kappa^2\right)+\sqrt{3}m\kappa(\dot{Q}+5HQ)+3\kappa^2 (\dot{H}+4H^2)\big]Q^2-V^\prime\nonumber\\
 &-\f92 \beta_1^\prime H (\beta_1 \dot{Q}+Q\beta_1^\prime\dot{\phi})Q-\f32 \beta_1^\prime (2\beta_1^\prime\dot{\phi}\dot{Q}+\beta_1 \ddot{Q}+Q\beta_1^\prime\ddot{\phi}+Q\beta_1^{\prime\prime}\dot{\phi}^2)Q-\ddot{\phi}=0.
\end{align}
The $(0,0)$ and $(1,1)$ components of the metric equation can be written as
\begin{align}
3\kappa^2 H^2 &-\f12 V-\f{m^2}{8\kappa^2}  \left(2\kappa^2(1-\beta_1^2)+3\beta_1^2 Q^2\right)Q^2-\f38 (\beta_1 \dot{Q}-Q \dot{\beta_1})^2+\f34\beta_1 (\beta_1 \ddot{Q}+Q \ddot{\beta_1})Q\nonumber\\
&-\f{\sqrt{3}m}{4\kappa}\beta_1 (6H\beta_1- \dot{\beta_1})Q^3+3H\beta_1 (\beta_1 \dot{Q}+Q \dot{\beta_1})Q-\f38 \beta_1^2 (9H^2+2\dot{H})Q^2-\f14\dot{\phi}^2 =0,
\end{align}
and
\begin{align}
\f18\big(8\kappa^2&-\beta_1^2 Q^2\big)(3H^2+2\dot{H})-\f12 V+\f{m^2}{8\kappa^2} \left(2\kappa^2(1-\beta_1^2)+\beta_1^2 Q^2\right)Q^2+ \f58 (\beta_1\dot{Q}+Q\dot{\beta_1})^2+\f14\dot{\phi}^2\nonumber\\&-\f{\sqrt{3}m}{12\kappa}\beta_1(6\beta_1 \dot{Q}+5 Q \dot{\beta_1}) Q^2 -\f12 H\beta_1 (\beta_1 \dot{Q}+Q\dot{\beta_1})Q +\f14 \beta_1 (2\dot{\beta_1}\dot{Q}+\beta_1 \ddot{Q}+Q \ddot{\beta_1})Q=0,
\end{align}
respectively. In the above equations prime denotes derivative with respect to the argument and dot denotes derivative with respect to time. Note that $\beta_2$ does not appear in the background cosmological equations.

Let us now assume that the potential term has a form $V=\lambda\phi^2$ and the pseudo scalar function $\beta_1$ takes the simplest form $\beta_1=\phi$. The implications of other types of potential terms in the evolution of the universe is briefly discussed in \cite{mg14}. For the dS solution the Hubble parameter is constant $H=H_0$ and we also assume that $Q=Q_0$‌ and $\phi=\phi_0$ are also constants. The cosmological equations will then be reduced to
\begin{align}
&\phi_0^2 (Q_1^2+3 \sqrt{3} Q_1 H_1+6
H_1^2-1)+1=0,\nonumber\\
&Q_1^2 (Q_1^2+5 \sqrt{3} Q_1 H_1+12
H_1^2-2)-4 \Lambda=0,\nonumber\\
&3 Q_1^4 \phi_0^2+12 \sqrt{3} Q_1^3 H_1
\phi_0^2+Q_1^2 \left(\left(27 H_1^2-2\right)
\phi_0^2+2\right)-24 H_1^2+4 \Lambda  \phi_0^2=0,\nonumber\\
&Q_1^2 \left(\phi_0^2 \left(-Q_1^2+3
H_1^2+2\right)-2\right)-24 H_1^2+4 \Lambda  \phi_0^2=0.
\end{align}
where we have defined dimensionless quantities as
\begin{align}
Q_1=\f{Q_0}{\kappa},\qquad H_1=\f{H_0}{m},\qquad \Lambda=\f{\lambda}{m^2\kappa^2}.
\end{align}
One can check that the above system has a solution
\begin{align}\label{ds}
H_1=0.83,\qquad \phi_0^2=5.84,\qquad Q_1=-1.00, \qquad \Lambda=0.02.
\end{align}
In the next section we will take this as a background dS solution of the theory.
\section{Cosmological perturbations around de Sitter background}\label{sec5}
In this section we will perform the cosmological perturbation analysis around the de Sitter solution \eqref{ds} obtained in the previous section. For the metric perturbation around flat FRW background, we assume that the line element can be written as
\begin{align}\label{100}
ds^2=-(1+2\varphi)\,dt^2+2a(S_i+\partial_i B)dx^i\, dt+a^2\big((1+2\psi)\delta_{ij}+\partial_i\partial_j E+\partial_{(i}F_{j)}+h_{ij}%
\big)dx^i dx^j,
\end{align}
where $\varphi$, $\psi$, $E$ and $B$ are the scalar perturbations, $S_i$ and $F_i$ are the vector perturbations with vanishing divergence $\partial_i S_i=0=\partial_i F_i$, and $h_{ij}$ is the traceless and transverse tensor perturbation,
 $h_{ii}=0=\partial_i h_{ij}$. The spatial indices are raised and lowered by $\delta_{ij}$. We decompose the vector field as
\begin{align}\label{101}
Q_\mu=(Q_0+\delta Q_0,\xi_i+\partial_i \delta Q),
\end{align}
where $Q_0$ is the background value, $\delta Q_0$ and $\delta Q$ are the scalar perturbations and $\xi_i$ is a transverse vector perturbation $\partial_i\xi_i=0$. The axion field can also be decomposed as
\begin{align}\label{102}
\phi=\phi_0+\delta\phi.
\end{align}
The system then has two tensor dof associate with $h_{ij}$, six vector dof and seven scalar dof in total. In the perturbation analysis of the theory it will be easier if one write the perturbed action in terms of the gauge invariant quantities. For this, one should mention that under the infinitesimal coordinate transformations of the form $x^\mu\rightarrow x^\mu+\delta x^\mu$, the scalar perturbations transform as
	\begin{align}
	&\varphi\rightarrow\varphi-\partial_t\delta x^0,\qquad\quad
	B\rightarrow B+\f{1}{a}\delta x^0-a\partial_t\delta x,\qquad
	\psi\rightarrow\psi-H\delta x^0,\qquad
	E\rightarrow E-2\delta x,\nonumber\\
	&\delta Q\rightarrow \delta Q-Q_0\delta x^0,\quad~
	\delta Q_0\rightarrow \delta Q_0-Q_0\partial_t\delta x^0,\qquad~
~\delta\phi\rightarrow\delta\phi.
	\end{align}
The vector perturbations will transform as
\begin{subequations}
	\begin{align}
	S_i\rightarrow S_i-a\partial_t\eta_i,\qquad
	F_i\rightarrow F_i-2\eta_i,\qquad
	\xi_i\rightarrow\xi_i,
	\end{align}
\end{subequations}
and the tensor perturbation remains invariant under this transformation, $h_{ij}\rightarrow h_{ij}$. Note that we have decomposed the coordinate differentials as $\delta x^\mu=(\delta x^0,\delta^{ij}\big(\partial_j\delta x+\eta_j)\big)$.

 The background quantities $(Q_0,\phi_0,H_0)$ are all constant as obtained in the previous section, e.g. equation \eqref{ds}. 

From the above expressions, one can obtain five independent gauge invariant scalar perturbations, one of them is $\delta\phi$‌ and the others are
\begin{align}
\Phi&=\varphi+\partial_t\left(aB-\f{a^2}{2}\partial_t E\right),\qquad\qquad
\Psi=\psi+H\left(aB-\f{a^2}{2}\partial_t E\right),\\
\delta \mathcal{Q}_0&=\delta Q_0+Q_0\partial_t\left(aB-\f{a^2}{2}\partial_t E\right),\qquad
\delta \mathcal{Q}=\delta Q+Q_0\left(aB-\f{a^2}{2}\partial_t E\right),
\end{align}
And two independent gauge invariant vector perturbations
\begin{align}
\rho_i=S_i-\f12a\partial_t F_i,\qquad \xi_i\rightarrow\xi_i
\end{align}
Also, we have one independent gauge invariant tensor perturbation $h_{ij}$.

By substituting the perturbed quantities \eqref{100}-\eqref{102}, one can see that the scalar, vector and tensor parts of the action decompose from each other. As a result, in the following we will consider these perturbations separately.
\subsection{Tensor perturbations}
The tensor perturbation $h_{ij}$ has two polarization modes which can be represented as $h_{+}$ and $h_{\times}$. After Fourier transforming the perturbation fields, one can obtain the second order action of tensor perturbation as
\begin{align}
S^{(2)}_{tensor}=\f12\sum_{+,\times}\int\, d^3k\,dt\, \kappa^2\,a^3 A \left[\dot{h}_{ij}\dot{h}_{ij}-\f{k^2}{a^2}\f{1}{A}h_{ij}h_{ij}\right],
\end{align}
where we have defined $A=1+Q_1^2\phi_0^2$. Also $\kappa^2=1/16\pi G$ and $\vec{k}$ is the wave vector. One can see that the quantity $A$ is positive which implies that the kinetic term has a positive sign and there is no ghost and gradient instabilities in the tensor sector. In summary, we have two healthy tensor polarization degrees of freedom around de Sitter space-time without imposing any constraint on the parameters of the theory. 
\subsection{Vector perturbation}
For the vector sector of the theory, we have two gauge invariant vector perturbations $\rho_i$ and $\xi_i$. In terms of these quantities, one can write the vector sector of the perturbed action up to second order in perturbations as
\begin{align}\label{vecp1}
S^{(2)}_{vector}=&\int d^3kdt \Bigg[\left(\f12-\phi_0^2-\f{1}{\alpha}\beta_{20}+\f{1}{\alpha}\phi_0^2\beta_{20}\right)a\,\dot{\xi}_i^2-\f{k^2}{2\alpha a}\left(\alpha-2(\beta_{20}+\alpha)\phi_0^2+2\phi_0^4\right)\xi_i^2\nonumber\\&+k^2Q_0\phi_0^2\rho_i\xi_i+\f12k^2(\kappa^2+Q_0^2\phi_0^2)\,a\,\rho_i^2-\f{2\phi_0}{\alpha a^3}(1+2\alpha-4\beta_{20})\,\vec{k}\cdot(\vec{\xi}(k)\times\vec{\dot{\xi}}(-k))\Bigg],
\end{align}
where we have defined $F_{ij}=\partial_i\xi_j-\partial_j\xi_i$ ad. 
One can see from the action \eqref{vecp1} that the field $\rho_i$ is non-dynamical. Varying the above action with respect to $\rho_i$, gives
\begin{align}
\rho_i=-\f{Q_0\phi_0^2}{(\kappa^2+Q_0^2\phi_0^2)a}\xi_i.
\end{align}
Substituting for $\rho_i$ in \eqref{vecp1}, one can obtain a second order perturbed action for $\xi_i$
\begin{align}
S^{(2)}_{vector}=\f12\int d^3k\,dt \, a\, \Bigg[B_1\dot\xi_i^2-\f{k^2}{a^2}\left(B_1+\f{Q_0^2\phi_0^4}{\kappa^2+Q_0^2\phi_0^2}\right)\xi_i^2+\f{B_2}{a^4}\,\vec{k}\cdot(\vec{\xi}(k)\times\vec{\dot{\xi}}(-k))\Bigg],
\end{align} 
where we have defined
\begin{align}
B_1=1-2\phi_0^2-\f{2}{\alpha}\beta_{20}+\f{2}{\alpha}\phi_0^2\beta_{20},\qquad B_2=-\f{2\phi_0}{\alpha}(1+2\alpha-4\beta_{20}).
\end{align}
One can see from the above action that the theory has two propagating vector degrees of freedom associated with the vector field $Q_\mu$. In order to have an instability and ghost free theory, one should have $B_1>0$. Using equation \eqref{ds} and assuming $\beta_2=\phi^2$ for simplicity, one can see that $\alpha$ should satisfy the condition 
\begin{align}
\label{cond}
0<\alpha<\f{2\phi_0^2(1-\phi_0^2)}{1-2\phi_0^2}=5.3.
\end{align}
%In this case, we have two healthy propagating degree of freedom. In a special case $\alpha=2\beta_{20}(1-\phi_0^2)/(1-2\phi_0^2)$, the quantity $\xi_i$ will become non-dynamical with equation
%\textcolor{red}{$$\vec{\xi}=-\f{3H_0B_2(\kappa^2+Q_0^2\phi_0^2)}{2Q_0^2\phi_0^2k^2 a^2}\, (\vec{\xi}\times \vec{k}),$$}
%and we have no vector sector in the theory.
\subsection{Scalar perturbation}
Let us now consider the scalar sector of the theory. For simplicity in the following, we will assume that $\beta_2=\phi^2$. As was discussed before, there are 5 gauge invariant scalar perturbations, which can be written collectively as
$\mathcal{X}^T=(\delta\mathcal{Q},\mathcal{H},\Psi,\delta\phi,\Phi)$, where $\mathcal{H}\equiv Q_0^{-1}\delta\mathcal{Q}_0$ is the dimensionless helicity-0 perturbation of the vector field. The second order action of the scalar sector reduces to
\begin{align}\label{scl}
S^{(2)}_{scalar}&=\f12\int\, d^3k\,dt\, a^3\Bigg[3Q_0^2\Big(\f{k^2}{a^2}\dot{\mathcal{X}}^T\mathcal{K}_1\dot{\mathcal{X}}+\dot{\mathcal{X}}^T\mathcal{K}_2\dot{\mathcal{X}}\Big)\nonumber\\&+\Big(\f{k^4}{a^4}\mathcal{X}^T\mathcal{M}_1\mathcal{X}+\f{k^2}{a^2}\mathcal{X}^T\mathcal{M}_2\mathcal{X}+\mathcal{X}^T\mathcal{M}_3\mathcal{X}\Big)+\Big(\f{k^2}{a^2}\mathcal{X}^T\mathcal{R}_1\dot{\mathcal{X}}+\mathcal{X}^T\mathcal{R}_2\dot{\mathcal{X}}\Big) \Bigg],
\end{align}
where the matrices have defined in the appendix \ref{app1}. By calculating the  eigenvalues of the Kinetic matrix and using equation \eqref{ds}, one obtains
\begin{align}
0,\quad \f{m^2k^2}{\kappa^2a^2}\left(\f{37.65}{\alpha}-7.10\right),\quad 0.61,\quad -6.06,\quad 16.70,
\end{align}
showing that we have 4 dynamical scalar degrees of freedom in the theory, at least one of them is unstable (note that in the case $0<\alpha<5.30$ only one mode is unstable). For the sake of simplicity of calculations, let us consider from now on the $k\rightarrow\infty$ which corresponds to the deep inside horizon limit. In this case, the kinetic matrix $\mathcal{K}_2$ can be omitted compared to $\mathcal{K}_1$ and theory will have one scalar degree of freedom which is $\delta\mathcal{Q}$. Also, in this limit $\mathcal{M}_3$ and $\mathcal{R}_2$ can be dropped from the action \eqref{scl}. After integrating by part and simplifying the result, one obtains the second order action of the scalar perturbation as
\begin{align}
S^{(2)}_{scalar}&=\f12\int\, d^3k\,dt\,a^3\,m^4\f{k_1^2}{a^2}\left[A\,\dot{\delta\mathcal{Q}}^2+\left(B_1+\f{k_1^2}{a^2}B_2\right)\delta\mathcal{Q}^2\right],
\end{align}
where we have defined dimensionless wave vector $\vec{k}_1$ as $m\vec{k}_1=\vec{k}$ and
\begin{align}
A&=\frac{0.57 \alpha ^2-1.41 \alpha-9.01}{ (\alpha +3.03)
	(\alpha -9.96)},\\
B_1&=\frac{1.26 \alpha ^2+2.63 \alpha +1.26}{ (\alpha +3.03)
	(\alpha -9.96)},\\
B_2&=\frac{8.77 \alpha ^2-60.78 \alpha -265.02}{(\alpha +3.03)
	(\alpha -9.96)}.
\end{align}
In the above expression we have used the dS solution \eqref{ds} to simplify the result. The no-ghost and Tachyon condition will be obtained by imposing $A,B_1,B2>0$, which translates to
\begin{align}
\alpha <-3.03\quad \textmd{or}\quad -1.315<\alpha <3.41\quad \textmd{or}\quad \alpha >9.96.
\end{align}
Combining the above condition with \eqref{cond} one can deduce that the parameter $\alpha$ should be restricted in the region
\begin{align}
0<\alpha<3.41,
\end{align}
in order to have a healthy vector and scalar perturbations (at least in deep inside the horizon limit).
\section{Conclusions}\label{sec6}
In this paper, we have considered a Gauss-Bonnet action in Cartan space-time. In Cartan geometry, the torsion tensor is non-zero and the geometry of the space-time is determined by the metric and the torsion tensor. The torsion tensor in general has three independent part, one of them is related to the trace part and the other is related to the axial part of the torsion tensor.‌ The rest components can be described by a traceless tensor $t_{\mu\nu\rho}$. In this paper, we have obtained that the structure of the trace part of torsion tensor in Gauss-Bonnet Lagrangian resembles the vector Galileon Lagrangian. This vector Galileon term however, can be reduced to the Proca term after integration by parts. This suggests that the trace part of the torsion tensor in Gauss-Bonnet gravity produces a healthy second order theory. The possibility of producing healthy higher order vector theories from the trace part of the torsion tensor lies on the consideration of higher order Lovelock invariants in Cartan theory which will be the scope of the future works. In this work we have investigated the role of axial part of the torsion tensor in gravitational theory. The most interesting term of the axial vector field in Gauss-Bonnet theory is $\beta\epsilon^{\alpha\beta\gamma\delta}Q_{\alpha\beta}S_{\gamma\delta}$ which is total derivative unless $\beta$‌ is a scalar field. In this regards, we have promoted the parameter $\beta$ in the Gauss-Bonnet action to a scalar field.

The theory has a de Sitter expanding solution with healthy tensor and vector fluctuations. The tensor mode remains stable for all values of the model parameter, but the vector sector put a constraint on $\alpha$. The scalar perturbation however, contains four propagating modes. One of them is always unstable and the other two are always stable. In the case $0<\alpha<5.30$ the remaining degree of freedom is also stable. In this paper, we have analyzed the scalar sector in the deep inside the horizon limit when $k\rightarrow\infty$. In this regime, only one of the scalar modes remains dynamical and it is also healthy provided that the constraint on the vector perturbation holds. In this limit, we have left with a massless tensor mode, a massless vector mode and a scalar mode.
\subsection*{Acknowledgements}
We would like to thank the anonymous referee for very useful comments.
\appendix
\section{Scalar perturbation matrices}\label{app1}
The kinetic matrices can be written as
\begin{align}
\mathcal{K}_1=\left(
\begin{array}{ccccc}
\frac{2 A_1}{3 Q_0^2} & 0 & 0 & 0 & 0 \\
0 & 0 & 0 & 0 & 0 \\
0 & 0 & 0 & 0 & 0 \\
0 & 0 & 0 & 0& 0 \\
0 & 0 & 0 & 0 & 0 \\
\end{array}
\right),\qquad \mathcal{K}_2=\left(
\begin{array}{ccccc}
0 & 0 & 0 & 0 & 0 \\
0 & \phi _0^2 & -\phi _0^2 & \phi _0 & -\phi _0^2 \\
0 & -\phi _0^2 & (-\f{8\kappa^2}{Q_0^2}+\phi_0^2) & -\phi _0 & \phi _0^2 \\
0 & \phi _0 & -\phi _0 & 1+\f{2\kappa^2}{3Q_0^2	} & -\phi _0 \\
0 & -\phi _0^2 & \phi _0^2 & -\phi _0 & \phi _0^2 \\
\end{array}
\right).
\end{align}
The mass and gradient matrices are
\begin{align}
\mathcal{M}_1=\f32\phi_0^2\left(
\begin{array}{ccccc}
1 & 0 & 0 & 0 & 0 \\
0 & 0 & 0 & 0 & 0 \\
0 & 0 & 0 & 0 & 0 \\
0 & 0 & 0 & 0 & 0 \\
0 & 0 & 0 & 0 & 0 \\
\end{array}
\right),\quad
\mathcal{M}_2=\f12\left(
\begin{array}{ccccc}
0 & A_8 \phi _0 & -H_0 Q_0 \phi _0^2 & A_{11} & A_{13} \\
\phi _0  A_8 & 2 Q_0^2A_1  & 0 & 2 A_3 & 4 Q_0^2 \phi _0^2
\\
-H_0 Q_0 \phi _0^2 & 0 & 8 \kappa ^2 & 0 & 8 \kappa ^2 \\
A_{11} & 2 A_3 & 0 & 2 (A_3-\kappa^2) & 4 Q_0^2 \phi _0 \\
A_{13} & 4 Q_0^2 \phi _0^2 & 8 \kappa ^2 & 4 Q_0^2 \phi _0 & -8 Q_0^2
\phi _0^2 \\
\end{array}
\right),\nonumber
\end{align}
\begin{align}
\mathcal{M}_3=\f14\left(
\begin{array}{ccccc}
0 & 0 & 0 & 0 & 0 \\
0 & 4 A_5 Q_0^2 & 2 A_{10} Q_0^2 & 2 A_7 Q_0^2 & 2
A_{12} Q_0^2 \\
0 & 2 A_{10} Q_0^2 & 4 A_2 Q_0^2 & -15 A_8 H_0 Q_0 & 2
A_9 Q_0^2 \\
0 & 2 A_7 Q_0^2 & -15 A_8 H_0 Q_0 & 0 & 2 A_{14} \\
0 & 2 A_{12} Q_0^2 & 2 A_9Q_0^2 & 2A_{14}& 4 A_6
Q_0^2 \phi _0^2 \\
\end{array}
\right),
\end{align}
and the mixed matrices are
\begin{align}
\mathcal{R}_1=\left(
\begin{array}{ccccc}
0& 0 & 0 & 0 & 0 \\
Q_0 A_{16} & 0 & 0 & 0 & 0 \\
-Q_0\phi_0^2 & 0 & 0 & 0 & 0 \\
\f{Q_0\phi_0}{\alpha}(\alpha-4\phi_0^2) & 0 & 0 & 0 & 0 \\
3Q_0\phi_0^2 & 0 & 0 & 0 & 0 \\
\end{array}
\right),\quad
\mathcal{R}_2=\left(
\begin{array}{ccccc}
0 & 0 & 0 & 0 & 0 \\
0 & 0 & 0 & \f{\sqrt{3}m}{\kappa}Q_0^3\phi_0 & -3H_0Q_0^2\phi_0^2 \\
0 & \f{Q_0^2}{3H_0}A_{10} & 0 & \f52Q_0A_8 & A_15 \\
0 & \f{2\sqrt{3}m}{\kappa}Q_0^3\phi_0 & 0 & 0 & Q_0A_{17} \\
0 & 0 & 0 & -\f{\sqrt{3}m}{\kappa}Q_0^3\phi_0 & 0 \\
\end{array}
\right),
\end{align}
and we have defined the coefficients as
\begin{align}
A_1&=\frac{1}{\alpha}\big(\alpha-2 (\alpha +1) \phi _0^2 +2 \phi _0^4\big),\quad
A_2=\frac{3 \sqrt{3} H_0 m Q_0 \phi _0^2}{\kappa }+9 H_0^2 \phi _0^2-\frac{3}{2}
m^2 \phi _0^2-\frac{3 m^2}{2}-\frac{6 \lambda  \phi _0^2}{Q_0^2},\nonumber\\
A_3&=\frac{2 Q_0^2 }{\alpha }\left(\phi _0^2-\alpha \right),\quad
A_4=\frac{3}{2} Q_0^2 \phi _0^2-12 \kappa ^2,\quad
A_5=\frac{m Q_0 \phi _0^2 }{\kappa ^2}\left(3 \sqrt{3} H_0 \kappa +2 m Q_0\right),\nonumber\\
A_6&=\frac{-72 H_0^2 \kappa ^4+3 Q_0^2 \left(45 H_0^2 \kappa ^2+20 \sqrt{3} H_0
	\kappa  m Q_0+m^2 \left(5 Q_0^2-2 \kappa ^2\right)\right)+4 \kappa ^2
	\lambda +6 \kappa ^2 m^2 Q_0^2}{4 \kappa ^2 Q_0^2},\nonumber\\
A_7&=\frac{18 \sqrt{3} H_0 m Q_0 \phi _0}{\kappa }+24 H_0^2 \phi _0+\frac{4 m^2
	Q_0^2 \phi _0}{\kappa ^2}-4 m^2 \phi _0,\quad
A_8=\frac{2 Q_0 \phi _0 }{\kappa }\left(3 H_0 \kappa +\sqrt{3} m Q_0\right),\nonumber\\
A_9&=-\frac{36 H_0^2 \kappa ^2}{Q_0^2}+\frac{27}{2} H_0^2 \phi _0^2-\frac{9 m^2
	Q_0^2 \phi _0^2}{2 \kappa ^2}+3 m^2 \phi _0^2-3 m^2-\frac{6 \lambda  \phi
	_0^2}{Q_0^2},\nonumber\\
A_{10}&=-\frac{9 H_0 \phi _0^2 }{\kappa }\left(5 H_0 \kappa +2 \sqrt{3} m Q_0\right),\quad
A_{11}=6 H_0 Q_0 \phi _0+\frac{5 m Q_0^2 \phi _0}{\sqrt{3} \kappa },\nonumber\\
A_{12}&=-\frac{12 \sqrt{3} H_0 m Q_0 \phi _0^2}{\kappa }-24 H_0^2 \phi _0^2-\frac{4
	m^2 Q_0^2 \phi _0^2}{\kappa ^2},\quad
A_{13}=-7 H_0 Q_0 \phi _0^2-\frac{2 \sqrt{3} m Q_0^2 \phi _0^2}{\kappa },\nonumber\\
A_{14}&=-\frac{18 \sqrt{3} H_0 m Q_0^3 \phi _0}{\kappa }-36 H_0^2 Q_0^2 \phi _0-4
\lambda  \phi _0-\frac{3 m^2 Q_0^4 \phi _0}{\kappa ^2}+2 m^2 Q_0^2 \phi _0,\nonumber\\
A_{15}&=-24 H_0 \kappa ^2+18 H_0 Q_0^2 \phi _0^2+\frac{6 \sqrt{3} m Q_0^3 \phi
	_0^2}{\kappa },\quad
A_{16}=-2A_1-3\phi_0^2,\nonumber\\
A_{17}&=-3 H_0 Q_0 \phi _0-\frac{2 \sqrt{3} m Q_0^2 \phi _0}{\kappa }.
\end{align}


\begin{thebibliography}{99}
		\bibitem{accel} A. G. Riess et al., Astron. J. 116, 1009 (1998); S. Perlmutter et al., Astrophys. J. 517, 565 (1999); R. A. Knop et al.,
		Astrophys. J. 598, 102 (2003); R. Amanullah et al., Astrophys. J. 716, 712 (2010).
		\bibitem{LCDM} Planck Collaboration: P. A. R. Ade et al., Astron. Astrophys. 594, A13 (2016).
		\bibitem{CCproblem}  S. Weinberg, Rev. Mod. Phys. 61, 1 (1989).
		\bibitem{modi}T. Clifton, P. G. Ferreira, A. Padilla and C. Skordis, Phys. Rep. 513, 1 (2012).
		\bibitem{scalten} Y. Fujii and K. Maeda, Class. Quantum Grav. 20, 4503 (2003); T. Singh and T. Singh, Int. J. Mod. Phys. A 2, 645 (1987).
		\bibitem{vecten} G. Esposito-Farese, C. Pitrou and J. Uzan, Phys. Rev. D 81, 063519 (2010).
		\bibitem{galile} A. Nicolis, R. Rattazzi and E. Trincherini, Phys. Rev. D 79, 064036 (2009).
		\bibitem{DGP} G. Dvali, G. Gabadadze and M. Porrati, Phys. Lett. B 485, 208 (2000).
		\bibitem{super} P. de Fromont, C. de Rham, L. Heisenberg and A. Matas, JHEP 2013, 67 (2013); Heisenberg L. (2015) Superluminal Propagation in Galileon Models. In: Theoretical and Observational Consistency of Massive Gravity. Springer Theses (Recognizing Outstanding Ph.D. Research). Springer, Cham; K. Hinterbichler, A. Nicolis, and M. Porrati, JHEP 0909, 089 (2009); C. de Rham, L. Keltner, A. J. Tolley, Phys. Rev. D 90, 024050 (2014).
		\bibitem{nonrenor} G. Goon, K. Hinterbichler, A. Joyce and M. Trodden, JHEP 11,100 (2016); A. Nicolis and R. Rattazzi, JHEP 0406, 059 (2004); C. de Rham, G. Gabadadze, L. Heisenberg and D. Pirtskhalava, Phys. Rev. D 87, 085017 (2013).
		\bibitem{covgali} C. Deffayet, G. Esposito-Farese and A. Vikman, Phys. Rev. D 79, 084003 (2009).
		\bibitem{horn}  G.W. Horndeski, Int. J. Theor. Phys. 10, 363 (1974); M. Crisostomi, K. Koyama and G. Tasinato, JCAP 04, 044 (2016).
		\bibitem{appgali} A. De Felice and S. Tsujikawa, Phys. Rev. Lett. 105, 111301 (2010); T. Kobayashi, Phys. Rev. D 81, 103533 (2010); F. P Silva and K. Koyama,  Phys. Rev. D 80, 121301 (2009); N. Chow and J. Khoury, Phys. Rev. D 80, 024037 (2009);C. Deffayet, S. Deser and G. Esposito-Farese, Phys. Rev. D 80, 064015 (2009); A. De Felice, R. Kase and S. Tsujikawa, Phys. Rev. D 83, 043515 (2011); C. Burrage, C. de Rham, D. Seery and A. J. Tolley, JCAP 1101, 014 (2011); A. Padilla, P. M. Saffin and S. Zhou, JHEP 1012, 031 (2010); C. Deffayet, S. Deser and G. Esposito-Farese, Phys. Rev. D 82, 061501 (2010); C. Charmousis, E. J. Copeland, A. Padilla and P. M. Saffin, Phys. Rev. Lett. 108, 051101 (2012); E. Bellini, N. Bartolo and S. Matarrese, JCAP 1206, 019 (2012); A. De Felice, R. Kase and S. Tsujikawa, Phys. Rev. D 85, 044059  (2012); E. Babichev, Phys. Rev. D 86, 084037 (2012).
		\bibitem{nogot} C. Deffayet, A. E. Gumrukcuoglu, S. Mukohyama and Y. Wang, JHEP 04, 082 (2014).
		\bibitem{vecgali} L. Heisenberg, JCAP 05, 015 (2014).
		\bibitem{genervec} L. Heisenberg, arXiv:1801.01523 [gr-qc]; S. Nakamura, R. Kase and S. Tsujikawa, Phys. Rev. D 95, 104001 (2017); R. Kimura, A. Naruko and D. Yoshida, JCAP 1701, 002 (2017); L. Heisenberg, R. Kase and S. Tsujikawa, Phys. Lett. B 760, 617 (2016);J. B. Jimenez and L. Heisenberg, Phys. Lett. B 757, 405 (2016); M. Hull, K. Koyama and G.Tasinato, Phys. Rev. D 9, 064012 3 (2016); N. Khosravi, Phys. Rev. D 89, 124027 (2014); J. B. Jimenez, arXiv:1606.04361 [gr-qc]; Z. Haghani, T. Harko and S. Shahidi, Eur. Phys. J. C 77, 514 (2017); Z. Haghani, T. Harko‌ and S. Shahidi, arXiv:1707.00939v2 [gr-qc]; Z. Haghani, T. Harko, H. R. Sepangi and S. Shahidi, Eur. Phys. J. C 77, 137 (2017). 
		\bibitem{quantvec} F. Charmchi, Z. Haghani, S. Shahidi and L. Shahkarami, Phys. Rev. D 93, 124044 (2016); A. Amado, Z. Haghani, A. Mohammadi and S. Shahidi, Phys. Lett. B 772, 141 (2017).
		\bibitem{fR} A. De Felice and S. Tsujikawa, Living Rev. Rel. {\bf 13}, 3 (2010); T. P. Sotiriou and V. Faraoni, Rev. Mod. Phys. {\bf 82}, 451 (2010); S. Nojiri, S. D. Odintsov, Phys. Rept. 505, 59 (2011), arXiv:1011.0544; S. Nojiri, S.D. Odintsov and V.K. Oikonomou, Phys. Rept. 692, 1 (2017), arXiv:1705.11098.
		\bibitem{lovelock}D. Lovelock, J. Math. Phys. 12, 498 (1971); T. Padmanabhan and D. Kothawala, Phys. Rep. 531, 115 (2013); P. Bueno, P. A. Cano, O. Lasso A. and P. F. Ramirez, JHEP 1604, 028 (2016).
\bibitem{fRT} T. Harko, F. S. N. Lobo, S. Nojiri and S. D. Odintsov, Phys. Rev. D 84, 024020 (2011); T. Harko and F. S. N. Lobo, Eur. Phys. J. C 70, 373 (2010); F. S. N. Lobo, T. Harko, arXiv:1211.0426 [gr-qc]; Z. Haghani, T. Harko, F. S. N. Lobo, H. R. Sepangi and S. Shahidi, Phys. Rev. D 88, 044023 (2013); M. Roshan and F. Shojai, Phys. Rev. D 94, 044002 (2016); I. Ayuso, J. Beltran Jimenez and A. de la Cruz Dombriz, Phys. Rev. D 91, 104003 (2015) .
\bibitem{reunited} C. de Rham, A. J. Tolley, JCAP 1005, 015 (2010); K. Van Acoleyen and J. Van Doorsselaere, Phys. Rev. D 83, 084025 (2011).
\bibitem{massivereview}	K. Hinterbichler, Rev. Mod. Phys. 84, 671 (2012); C. de Rham, Living Rev. Relativity 17, 7 (2014).
\bibitem{FP} M. Fierz and W. Pauli, Proc. Roy. Soc. Lond., A173, 211 (1939).
\bibitem{dRGT}  C. de Rham and G. Gabadadze,
Phys. Rev., D 82, 044020 (2010); C. de Rham and G. Gabadadze, Phys. Lett. B. 693, 334 (2010); C. de Rham, G. Gabadadze and A. Tolley, Phys. Rev. Lett. 106, 231101 (2010); S. F. Hassan and R. A. Rosen, Phys. Rev. Lett. 108, 041101 (2012); S. F. Hassan and R. A. Rosen, JHEP 1204, 123 (2012); S. F. Hassan and R. A. Rosen, JHEP 1107, 009 (2011); S. F. Hassan, R. A. Rosen and A. Schmidt-May, JHEP 1202, 026 (2012).
\bibitem{massiveapp} C. de Rham and L. Heisenberg, Phys. Rev. D 84, 043503  (2011); Th. M. Nieuwenhuizen, Phys. Rev. D 84, 024038  (2011); D. Comelli, M. Crisostomi, F. Nesti and L. Pilo, Phys. Rev. D 85, 024044 (2012); S.F. Hassan and R. A. Rosen, JHEP 1202, 126  (2012);	G. D'Amico, C. de Rham, S. Dubovsky, G. Gabadadze, D. Pirtskhalava and  A.J. Tolley, Phys. Rev. D 84, 124046 (2011); N. Khosravi, H. R. Sepangi and S. Shahidi, Phys. Rev. D 86, 043517 (2012); N. Khosravi, N. Rahmanpour, H. R. Sepangi and S. Shahidi, Phys. Rev. D85,  024049 (2012); A. E. Gumrukcuoglu, C. Lin and S. Mukohyama, JCAP 11, 030 (2011); A. E. Gumrukcuoglu, C. Lin and S. Mukohyama, JCAP 1203, 006 (2012); K. Hinterbichler, R. A. Rosen, JHEP 1207, 047  (2012); K. Nomura and J. Soda, Phys. Rev. D 86, 084052   (2012); Q. Huang, Y. Piao and S. Zhou, Phys. Rev. D 86, 124014 (2012); G. D'Amico, G. Gabadadze, L. Hui and D. Pirtskhalava, Phys. Rev. D 87, 064037  (2013); S. Nojiri and S. D. Odintsov, Phys. Lett. B 716, 377  (2012); Z. Haghani, H. R. Sepangi and S. Shahidi, Phys. Rev. D 87, 124014 (2013).
\bibitem{cartan} E. Cartan, C. R. Acad. Sci. (Paris) 174, 593 (1922); F. W. Hehl, P. von der Heyde, G. D. Kerlick, and J. M. Nester, Rev. Mod. Phys. 48, 393 (1976); F. W. Hehl and B. K. Datta, J. Math. Phys. 12, 1334 (1971); A. Trautman, arXiv:gr-qc/0606062v1. 
\bibitem{tor} T. B. Vasilev, J. A. R. Cembranos, J. G. Valcarcel and P. Martín-Moruno, Eur. Phys. J. C 77, 755 (2017).
\bibitem{WCGB} 	Z. Haghani, N. Khosravi and S. Shahidi, Class. Quant. Grav. 32 (2015) 215016; J. Beltran Jimenez and T. S. Koivisto, Class. Quant. Grav. 31 (2014) 135002.
\bibitem{more} J. Beltran Jimenez, L. Heisenberg and T. S. Koivisto, JCAP04, 046 (2016); J. Beltrán Jiménez and T. S. Koivisto, Phys. Lett. B 756, 400 (2016). 

\bibitem{torsiondecom} J. D. McCrea, Class. Quant. Grav. 9 (1992) 553; F.W. Hehl, J.D. McCrea, E.W. Mielke and Y. Ne’eman, Phys. Rep. 258 (1995) 1.
\bibitem{supergravity}A. Candiello and K. Lechner, Nucl. Phys. B 412, 479 (1994).
	\bibitem{axion} D. J. E. Marsh, Phys. Rep. 643, 1 (2016) ; P. Sikivie, Lect. Notes Phys. 741, 19 (2008).
	\bibitem{mg14} S. Shahidi and Z. Haghani, The fourteenth Marcel Grossmann meeting proceedings, pp. 1319 (2017).
\end{thebibliography}
\end{document}